 \documentclass[12pt,preprint]{aastex}

\shorttitle{Lithium Abundances in Stars with Planets}
\shortauthors{Ghezzi et al.}

\begin{document}


\title{Lithium Abundances in a Sample of Planet Hosting Dwarfs\footnote{Based on observations 
made with the 2.2 m telescope at the European Southern Observatory (La Silla, Chile), 
under the agreement ESO-Observat\'orio Nacional/MCT.}}


\author{L. Ghezzi\altaffilmark{1}, K. Cunha\altaffilmark{1,2,3}, V. V. Smith\altaffilmark{2}, 
\& R. de la Reza\altaffilmark{1}}


\altaffiltext{1}{Observat\'orio Nacional, Rua General Jos\'e Cristino, 77, 20921-400, 
                 S\~ao Crist\'ov\~ao, Rio de Janeiro, RJ, Brazil; luan@on.br}
\altaffiltext{2}{National Optical Astronomy Observatory, 950 North Cherry Avenue, Tucson, AZ 85719, USA}
\altaffiltext{3}{Steward Observatory, University of Arizona, Tucson, AZ 85121, USA}


\begin{abstract}
This work presents a homogeneous determination of lithium 
abundances in a large sample of giant-planet hosting stars (N=117), 
and a control sample of disk stars without detected planets (N=145).
The lithium abundances were derived using a detailed profile fitting 
of the \ion{Li}{1} doublet at $\lambda$6708 \AA\ in LTE. 
The planet hosting and comparison stars were chosen to have significant overlap in their respective
physical properties, including effective temperatures,
luminosities, masses, metallicities and ages.
The combination of uniform data and homogeneous
analysis with well selected samples, makes this study well-suited
to probe for possible differences in the lithium abundances found
in planet hosting stars.
An overall comparison between the two samples reveals no
obvious differences between stars with and without planets. Closer
examination of the behavior of the Li abundances over a narrow range
of effective temperature (5700 K $\le$ T$_{\rm eff}$ $\le$ 5850 K)
indicates subtle differences between the two stellar samples; 
this temperature range is particularly sensitive to
various physical processes that can deplete lithium. 
In this T$_{\rm eff}$ range planet hosting stars have lower Li abundances (by $\sim$ 0.26 dex
on average) than the comparison stars, although this segregation
may be influenced by combining stars from a range of ages, metallicities
and masses. 
When stars with very restricted ranges in metallicity ([Fe/H] = 0.00 to +0.20 dex) and
mass (M $\sim$ 1.05 -- 1.15 M$_{\sun}$) are compared, however, both stars with
and without planets exhibit similar behaviors in the lithium abundance
with stellar age, suggesting that there are no differences in the lithium
abundances between stars with planets and stars not known to have planets.
\end{abstract}





\section{Introduction}

\label{int}

The physical processes that both create and destroy lithium lead to a complex behavior
of its chemical abundance in a variety of astrophysical sites.
The abundance of Li in stellar photospheres, in particular, holds important clues to the understanding of various processes, 
from Big Bang Nucleosynthesis to the formation and evolution of planetary systems (see, e.g., 
\citealt{m09}; \citealt{santos10}). The determination of lithium abundances in FGK dwarfs
is challenging as it requires high-resolution 
and high signal-to-noise (S/N) stellar spectra. In addition, the interpretation of the results is 
not straightforward as lithium abundances are known to depend on several variables, such as 
the effective temperature ($T_{eff}$), metallicity ([Fe/H]\footnote{Here, we adopt the usual notation 
[Fe/H] = $log(N_{Fe}/N_{H})_{\star} - log(N_{Fe}/N_{H})_{\sun}$}), stellar mass, age, rotation and activity level.

This complexity in the interpretation of lithium abundance results has led to conflicting 
conclusions concerning possible differences between lithium abundances in planet hosting stars
when compared with stars not known to host giant planets. 
In this introduction, we briefly discuss the results in these previous studies.
One of the first papers to focus on the element lithium in stars with planets was \citet{gl00}. 
This study compared Li abundances in a sample of 8 stars with planets to a sample of field stars 
taken from the literature. After correcting for systematic differences in $T_{eff}$, [Fe/H] and the
Ca II H+K emission index, $\log R'_{HK}$ 
(which is a measurement of the chromospheric emission), from the different studies, the conclusion was that stars 
with planets have lower Li abundances than non-planet hosts. 
This initial result, however, was not confirmed by \citet{r00}, who compared Li abundances (from the 
literature) of planet hosting stars with those of open cluster and field stars carefully selected to 
span similar intervals in temperature, age and chemical composition and found no differences in 
A(Li)\footnote {A(Li)= log[N(Li)/N(H)]+12}. 

As more planet hosting stars were discovered, the work by \citet{i04} obtained lithium abundances for a larger sample 
of 79 stars hosting planets and 38 stars without planets. 
This study did not find significant differences between the Li abundances in the two samples, but noted that 
the control sample had too few stars with detectable Li to be used as a comparison. 
The authors then compared their Li abundances for their sample of planet hosting stars with 
a sample of 157 field stars from \citet{c01} and concluded that 
planet hosting stars with effective temperatures in the range between 5600 K and 5850 K exhibit an excess of Li depletion. 
Two possible hypotheses were proposed for explaining this result: increased mixing due to rotational 
breaking caused by the interaction (during pre-main-sequence evolution) of the host stars with their 
proto-planetary disks; effective mixing generated by tidal forces resulting from planet migration. 

The findings of \citet{i04} were further confirmed by \citet{tk05}
with Li abundances derived for a sample of 160 FGK dwarfs and subgiants in the Galactic disk, among which 
there were only 27 planet hosting stars. The lithium abundance distributions for planet-hosting and non planet-hosting
stars in that study were found to be very similar; the size of 
the planet hosting stellar sample in \citet{tk05}, however, 
was not large enough in order to make meaningful statistical comparisons. 
If instead of using their own sample of planet-hosting stars, the authors adopted the Li results from \citet{i04},
for planet hosting stars only, a similar tendency of finding an excess of Li depletion in this sample
was obtained, although this was only observed for stars within the narrow range in effective temperature between 
5800 K and 5900 K. The study by \citet{cz06} analyzed smaller samples but 
also found a higher frequency of Li poor stars within their sample of 16 planet hosting stars in 
comparison with 20 control stars, but in this case for stars with effective temperatures ranging between 5600 K and 5900 K.

Contrary results to these findings were obtained by \citet{lh06}, continuing the controversy about Li
in planet-hosting stars. This study derived
Li abundances homogeneously for a sample of 216 nearby dwarf stars, 
55 of which host close giant planets. 
They found no differences between lithium abundances of stars with and without planets and argued
that the low-Li tendency observed by \citet{i04} was caused by a systematic difference between the temperature scales 
adopted in the latter work and the study of \citet{c01} (from which the control sample 
was drawn). 

The topic was revisited by \citet{g08} who combined
samples of stars from the literature (gathering a total of 37 stars with planets and 147 stars without planets)
and corrected for systematic offsets between the $T_{eff}$, log g, [Fe/H] and A(Li) 
from the different published analyses. \citet{g08} found that 
planet hosting stars with $T_{eff}$ $\sim$ 5800 K tend to have lower Li abundances compared to stars not
hosting planets. It was suggested in that study that stars with planets also exhibit larger Li abundances 
near $T_{eff} \simeq$ 6100 K, with the transition occurring at $T_{eff} \simeq$ 5950 K, 
and that $v \sin i$ and $\log R'_{HK}$ seem to correlate with the lithium abundance.

More recently, \citet{m09}, on the other hand, presented results 
for 4 and 6 stars with and without planets, respectively, 
with effective temperatures near T$_{eff}$ $\sim$ 5800 K (the same effective temperature regime 
for which \citealt{g08} finds differences).
Based on this much smaller but homogeneous set of results, \citet{m09} argue that stars with 
planets do not show anomalously low Li abundances.
A much larger sample observed with HARPS (GTO; \citealt{s08}) has been analyzed by \citet{i09}. 
Lithium abundances were obtained for 451 stars, among which there are 70 stars which host planets. This sample was 
extended to include 16 stars hosting planets and 13 stars without planets (with 5600 K $\lesssim 
T_{eff} \lesssim$ 5900 K). The results obtained from this sample confirmed the low-Li tendency for 
planet hosting stars with effective temperatures in the range between 5600 - 5900 K and 
$T_{eff} = 5777 \pm 80$ K (solar analogues). The question about differences in other parameters being
responsible for the effect was addressed by \citet{s10}, who showed that differences in 
mass and age could not be responsible for this enhanced depletion and proposed that the low lithium abundance 
is directly related to the presence of planets. 

Revisiting the topic, \citet{g10} 
performed a homogeneous analysis and derived Li abundances and $v \sin i$ for a 
large sample of stars with and without planets (with 90 and 60 stars, respectively). 
They basically confirm the results 
of \citet{g08}, although finding a lower temperature (T$_{eff}$ = 5850 K) for the transition region between
low and high Li abundances in stars with planets (relative to comparison stars). 
Finally, \citet{b10} analyzed a sample of 117 solar-like  stars, among which 
there are 14 planet-hosting stars that do not  exhibit an enhanced lithium depletion 
relative to stars without  planets. The differences between lithium abundances of stars 
with and without planets previously found in the literature were attributed to a 
systematic effect that arises from the comparison of samples with  different stellar 
properties, more specifically age and metallicity. 

From a theoretical point of view, recent calculations by \citet{bc10}
predict that episodic large accretion events from protoplanetary disks 
can enhance Li destruction in stars which host planetary systems. They suggest
that this is a possible mechanism for depleting lithium in stars with large
planets when compared to stars without.

\defcitealias{ghezzi10a}{Paper I}

In \citet[hereafter Paper I]{ghezzi10a}, we performed a homogeneous determination of stellar parameters and metallicities for 
a large sample of dwarf stars hosting giant planets close in, as well as a sample of control
disk stars not known to host large planets. The spectra analyzed in that study had high S/N and
covered the spectal region containing the \ion{Li}{1} feature at 6707.8 \AA.
The present study presents the homogeneous determination of lithium abundances for all the stars in \citetalias{ghezzi10a} based on a detailed 
profile fitting of the \ion{Li}{1} resonance doublet at $\lambda$6707.8 \AA. This paper is organized as 
follows: in Section \ref{data}, the sample and observational data are briefly described. The profile 
fitting techniques and resulting lithium abundances 
are presented in Section 
\ref{analysis}. Section \ref{disc} contains the discussion and interpretation of the results. 
Finally, concluding remarks are presented in Section \ref{conc}.


\section{Sample and Observational Data}

\label{data}

The sample of stars analyzed in this study is comprised of 117 stars hosting large planets and 145 comparison 
disk stars. The target stars were previously analyzed in \citetalias{ghezzi10a}, where stellar parameters and iron 
abundances were derived. 
The sample of main-sequence stars with planets were selected (until August 2008) from the 
Extrasolar Planet Encyclopaedia\footnote{Available at http://exoplanet.eu}
given the constraints on object observability from La Silla Observatory. 
In addition, a control sample of main-sequence stars without detected planets so far was compiled from 
the subset of 850 nearby FGK stars in \citet{fv05}, which has been monitored in planet search 
programmes. The selection criteria were such that the stellar properties of the control stars matched 
those of stars with planets (which makes this comparison sample adequate for the lithium study presented here). 

High-resolution (R $\sim$ 48,000) and high signal-to-noise (S/N $\sim$ 200) spectra were obtained with 
the Fiber-fed Extended Range Optical Spectrograph (FEROS; \citealt{kaufer99}) attached to the MPG/ESO-2.20m telescope 
(La Silla, Chile) during 6 observing runs between April 2007 and August 
2008\footnote{Under the agreement ESO-Observat\'orio Nacional/MCT}. The data reduction was done with the FEROS Data Reduction System 
(DRS)\footnote{Available at http://www.eso.org/sci/facilities/lasilla/instruments/feros/tools/DRS/index.html} 
and followed standard procedures for echelle spectra. 
A complete description of the sample selection, observations and data reduction can be found in \citetalias{ghezzi10a} 
and we refer to this previous study for details.


\section{Analysis}

\label{analysis}

The determination of Li abundances from synthetic spectra requires a line list for 
the spectral region around the \ion{Li}{1} feature at 6707.8 \AA. The line list adopted here was taken 
from \citet{ghezzi09}, but removing the $^{6}$Li components (not measurable at the resolution of FEROS spectra). 
The model atmospheres adopted in the calculations were interpolated from the ODFNEW grid of ATLAS9 
models\footnote{Available at http://kurucz.harvard.edu/} (\citealt{ck04}). Stellar parameters 
and metallicities ([Fe/H]) for the target stars were taken from Table 3 of \citetalias{ghezzi10a}. Briefly, the 
atmospheric parameters (T$_{eff}$, $\log g$ and $\xi$) were derived in Local Thermodynamic Equilibrium (LTE) 
following standard spectroscopic methods (excitation and ionization equilibria based on \ion{Fe}{1} and \ion{Fe}{2} lines;
see \citetalias{ghezzi10a} for all the details on these determinations). The 2002 version of the code MOOG\footnote{Available 
at http://www.as.utexas.edu/~chris/moog.html.} (\citealt{s73}) was used to compute synthetic spectra in
the Li region. The limb darkening coefficient was varied between zero and 1.0 (\citealt{vm93}) however,
the choice of coefficient has little, if any, effect on the \ion{Li}{1} line profile and the impact on the derived lithium 
abundances is negligible.

In this study, we adopted the following procedure in order to determine Li abundances for the target stars. 
First, it was assumed that the broadening of spectral lines was represented by a single Gaussian smoothing function 
which combined the effects of stellar rotation, macroturbulence and instrumental profile.
A grid of synthetic spectra was then computed 
for combinations of the Gaussian Full-Width at Half-Maximum (FWHM$_{Gauss}$) and lithium abundance. The best fit between 
the synthetic and observed profiles was obtained through a reduced $\chi^{2}$-minimization in the spectral interval between 
6707.3 - 6708.4 \AA\ (including mainly, besides the \ion{Li}{1} doublet, a Fe+CN and a Ca+CN feature). 
Small adjustments in the continuum level were allowed in order to compensate for possible errors in the 
continuum normalization. Also, wavelength shifts of the observed spectra were needed in order to account for 
radial velocity shifts. The abundances of Fe, CN, Si, Ca and V were 
kept as free parameters in this first step in order to properly match their contributions and improve the overall quality of the fits. 
Such adjustments in the elemental abundances, however, do not affect the derivation of the Li abundances, as their 
contributions to the global feature are clearly distinguishable from the \ion{Li}{1} line.
This first step provided the best-fit lithium abundance, as well as 
values of \textit{r} (continuum displacement), \textit{w} (wavelength shift) and FWHM$_{Gauss}$.
Such quantities were then kept fixed but 
the broadening was then separated into the contributions of macroturbulence and 
instrumental profile (combined as a single gaussian)
and the stellar rotational velocity profile. A new grid of synthetic spectra was computed.
The best fit between synthetic and observed profiles was again obtained through a 
reduced $\chi^{2}$-minimization in the spectral interval between 6707.3 - 6708.4 \AA. 
Given the intrinsic spectrograph resolution of FWHM$_{Inst}$ = 6.25 km s$^{-1}$, with typical macroturbulent
velocities of $\sim$ 4.00 km s$^{-1}$, the combined braodening is equivalent to 7.40 km s$^{-1}$. Tests reveal that
for the majority of stars sampled, which rotate rather slowly,
this analysis technique cannot distinguish between macroturbulence velocities and
projected rotational velocities $v \sin i$; because of the large relative uncertainty
in almost all of the values of v sin i in this way, we do not include any
further discussion of v sin i values determined here.

\begin{figure}
\epsscale{1.00}
\plotone{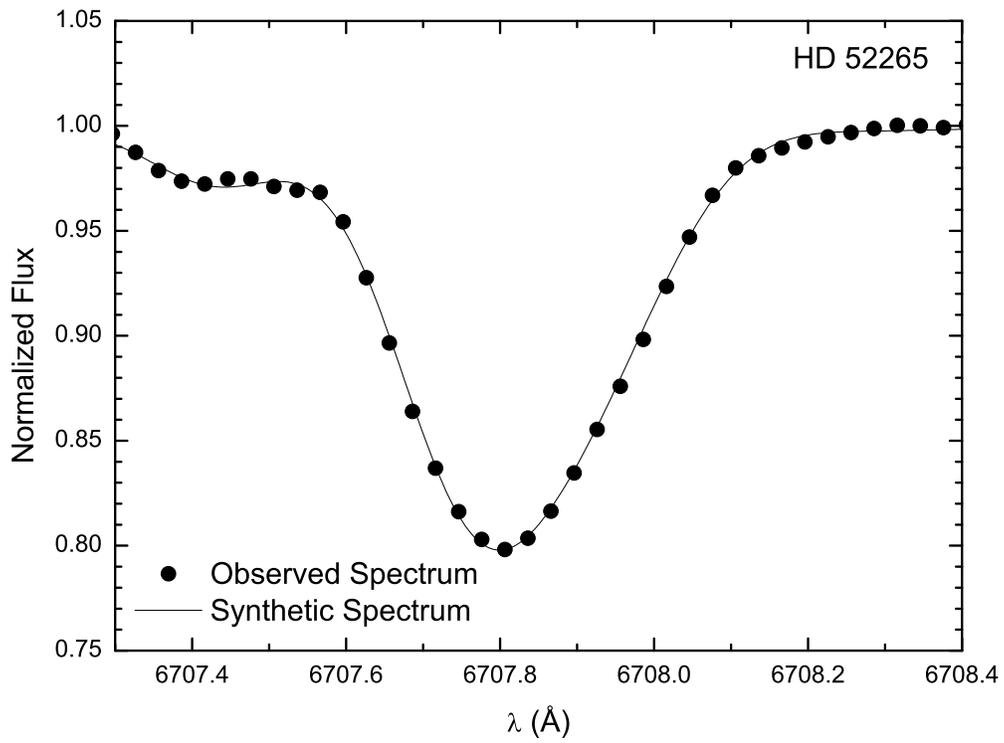}
\caption{The best fit obtained between observed (filled circles) and synthetic (solid line) 
spectra for target star HD 52265. 
The best-fit Li abundance displayed corresponds to A(Li) = 2.65.}
\label{fit}
\end{figure}

The determination of Li abundances described above was done in an automated way with the aid of BASH 
and FORTRAN codes: lithium abundances for the entire sample of 262 stars studied here could be derived in $\sim$2 days.
All automatically obtained fits were inspected visually and manual corrections were needed 
for approximately 10\% of the cases. The stars with non-optimum fits generally had spectra with low S/N values, 
or extremely weak or non-measurable Li lines for which only upper limits to the Li abundances could be derived. 
Figure \ref{fit} shows the best fit between observed and synthetic spectrum obtained for target star HD 52265 as an example. 
Table \ref{li} lists the values of FWHM$_{Gauss}$ (column 2) which include instrumental profile,
macroturbulence and stellar rotation, 
and derived A(Li) (column 3) for all studied stars.
The Li abundance in the Sun was also derived using a solar spectrum observed with the FEROS 
spectrograph on August 20, 2008. The result obtained for the Sun is A(Li) = 0.99 $\pm$ 0.16, which is in agreement with 
recent solar abundance determinations which are based on 3D hydrodynamical models and take non-LTE effects 
into account (A(Li) = 1.05 $\pm$ 0.10 from \citealt{a09}; A(Li) = 1.03 $\pm$ 0.03 from \citealt{c10}). 


We compared the results obtained here with those from \citet{ghezzi09}, 
which were based on much higher quality bHROS spectra (R $\sim$ 150,000 and S/N $\gtrsim$ 700). For the six 
stars in common (namely HD 17051, HD 36435, HD 74156, HD 82943, HD 147513 and HD 217107), 
we find the average difference (in the sense \textquotedblleft FEROS - bHROS\textquotedblright\ spectra) to be
$\langle\Delta$A(Li)$\rangle$ = 0.01 $\pm$ 0.09. The agreement between these two sets of results is very good.

\subsection{Uncertainties}

\label{error}

The formal uncertainties in the derived best-fit lithium abundances can be 
calculated by varying A(Li) around its best value and computing, for each lithium 
abundance tested, the quantity $\Delta\chi_{\rm r}^{2}$ = $\chi_{\rm r}^{2}$ - $\chi_{\rm r,min}^{2}$. The 
difference between A(Li) and A(Li)$_{best}$ that gives $\Delta\chi_{\rm r}^{2}$ = 1 is taken as the 1$\sigma$ 
uncertainty. 
The derived lithium abundances and their formal uncertainties are presented in column 3 of Table \ref{li}. 
The lithium abundances, however, are also sensitive to uncertainties in the effective temperature parameter: 
$\delta T_{eff}$ = $\pm$100 K introduces an uncertainty of $\sim$ $\pm$ 0.1 dex in the derived 
lithium abundances for most sample stars; for the cooler stars ($T_{eff} \sim$ 5000 K)
the sensitivity to $T_{eff}$ is slightly larger ($\sim$ 0.15 dex).
The sensitivity of A(Li) to variations in the adopted log g's and microturbulent velocities
is only marginal: a change in log g of $\pm$0.5 dex changes the lithium abundance by 0.01-0.02 dex,
with a slightly larger sensitivity of $\sim$ 0.05 for the coolest stars in our sample. A change in the
microturbulent velocity by $\pm$0.5 km s$^{-1}$ has virtually no effect on the derived A(Li).
The total uncertainties in the derived lithium abundances are then calculated by the addition of all these uncertainties 
in quadrature, assuming the errors are independent. 
The total estimated uncertainties in the derived Li abundances in this study are in column 4 of Table \ref{li}. 

Non-LTE effects have not been considered in this Li abundance analysis and these will contribute to
the total error budget. \citet{tk05} calculated non-LTE Li abundances for their sample
of stars with effective temperatures ranging between 5000 K and 7000 K, and metallicities 
-1.0 $< [Fe/H] <$ +0.40 dex. The non-LTE corrections in that study were found to range between -0.1 $< \Delta <$ +0.1 dex. 
Stars in restricted ranges in effective temperatures and metallicities, however, have similar non-LTE corrections. 
The stars with detected \ion{Li}{1} lines in this study have effective temperatures roughly between 5700 K and 6200 K
(as will be shown in Figure \ref{li_teff}). For these hotter dwarfs, with A(Li) $<$ 3.00, the non-LTE corrections are 
found to be small in the recent study of \citet{lind09}.


\section{Discussion}

\label{disc}

\subsection{Comparisons within the Samples: Metallicities, Masses, Ages and Activity}

\label{comp}

\begin{figure}
\plotone{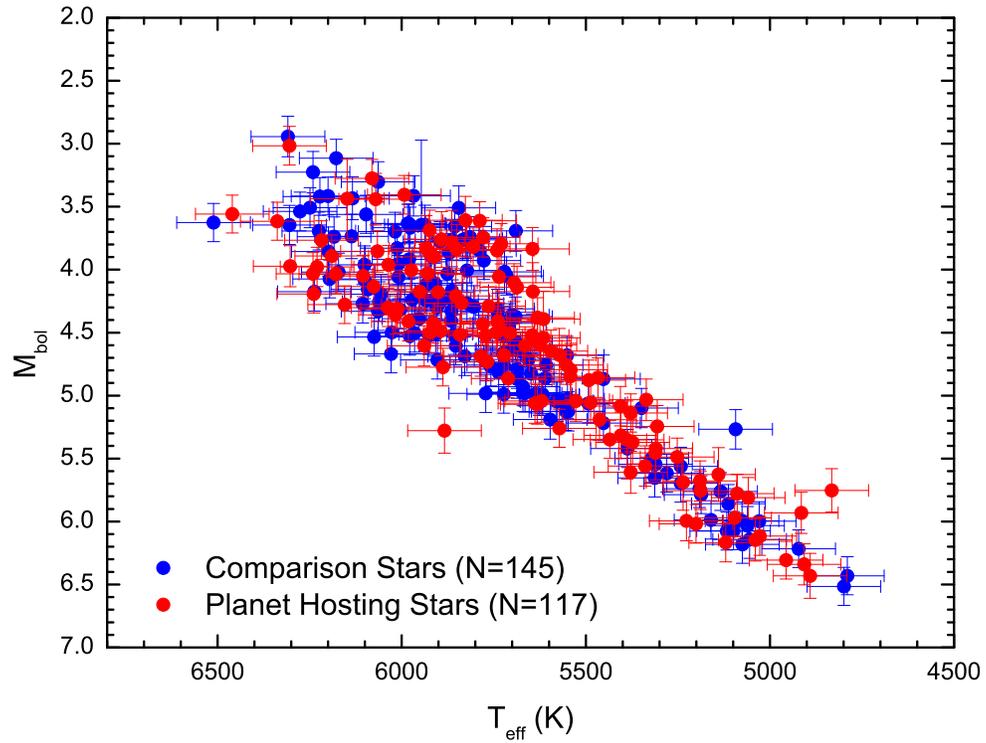}
\caption{Location of the target stars in an H-R diagram represented by the bolometric magnitudes and effective
temperatures. The planet hosting stars (red filled circles; N=117) and comparison stars not known to have
giant planets (blue filled circles; N=145) analyzed in this study are all unevolved.}
\label{Mbol_teff}
\end{figure}

As mentioned previously, interpreting lithium abundances can be a complex process 
because these abundances depend on a number of parameters, 
such as effective temperature, metallicity, mass, age, rotation, as well as the stellar activity level. The behavior of Li with 
$T_{eff}$ can only be isolated if samples with similar metallicities, masses, 
ages and activity levels are compared. The stars in the sample considered here are 
dwarfs, as shown in Figure \ref{Mbol_teff}; all targets lie on the main sequence.


\begin{figure}
\epsscale{1.00}
\plotone{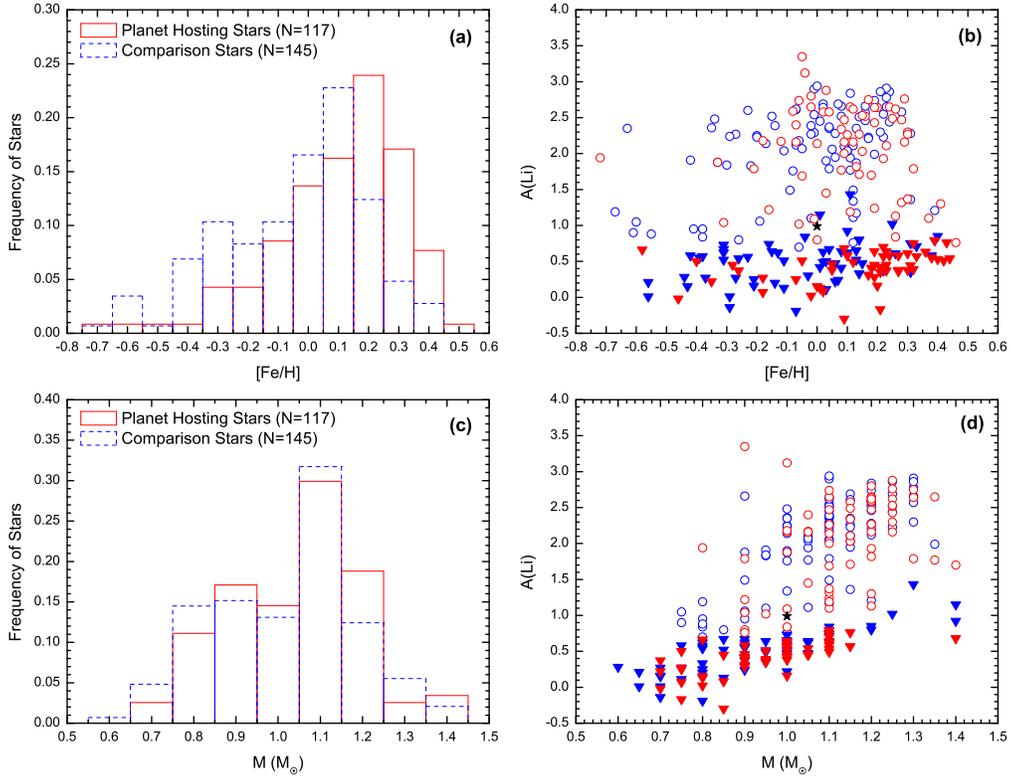}
\caption{(a) Metallicity distributions of planet hosting (solid red line) and comparison (blue dashed line) 
stars from \citetalias{ghezzi10a}. (b) Lithium abundances \textit{versus} metallicity for stars with (red symbols) and without (blue symbols)
planets. Open circles are detected abundances and inverted triangles denote upper limits. The Sun is represented 
by a black star. (c) Same as panel (a) but for masses. (d) Same as panel (b) but replacing metallicity with mass.}
\label{comp1}
\end{figure}

The metallicity distributions of the studied stars are shown in panel \textquoteleft a\textquoteright\ of Figure \ref{comp1}
(same as figure 9 of \citetalias{ghezzi10a}). As discussed in \citetalias{ghezzi10a} there is 
a visible offset between the distributions of stars with and without planets; the average metallicity of 
stars hosting giant planets is 0.15 dex higher than that of the comparison stars. 
This metallicity offset, however, has a negligible effect on the Li abundances, as can be noted from 
panel \textquoteleft b\textquoteright\ of Figure \ref{comp1} where Li is shown \textit{versus} [Fe/H]. 
The samples of stars with and without planets have considerable overlap with no significant trend of the lithium 
abundance with metallicity found over the roughly 1.0 dex change in [Fe/H] for the stars studied here. 
Also, the Sun does not exhibit any peculiarity in this plot.

Panel \textquoteleft c\textquoteright\ of Figure \ref{comp1} presents the mass distributions for the two studied samples. The adopted values for the stellar
masses correspond to the 
M$_{track}$ values presented in Table 4 of \citetalias{ghezzi10a} (see that paper for details on the derived masses). 
The samples of stars hosting planets (represented by the red solid line histogram) and without planets (represented by
the blue dashed line histogram) overlap closely in mass. 
The average masses for stars with and without planets are, respectively, $\langle M \rangle$ = 1.05 $\pm$ 0.16 and 
1.02 $\pm$ 0.17 M$_{\sun}$; there is a probability of 92\%, based on a Two-Sample Kolmogorov-Smirnov (KS) test 
that they are drawn from the 
same parent population of mass distributions. Panel \textquoteleft d\textquoteright\ of Figure \ref{comp1} reveals a tendency of higher Li abundances with 
increasing mass. This is expected (see, e.g., \citealt{lr04}), as lower mass stars have deeper 
convective zones in which Li is destroyed more efficiently. The location of the Sun in this diagram (represented
by the black star symbol) indicates
it to be a normal star when compared to the samples of stars with and without planets.


\begin{figure}
\epsscale{1.00}
\plotone{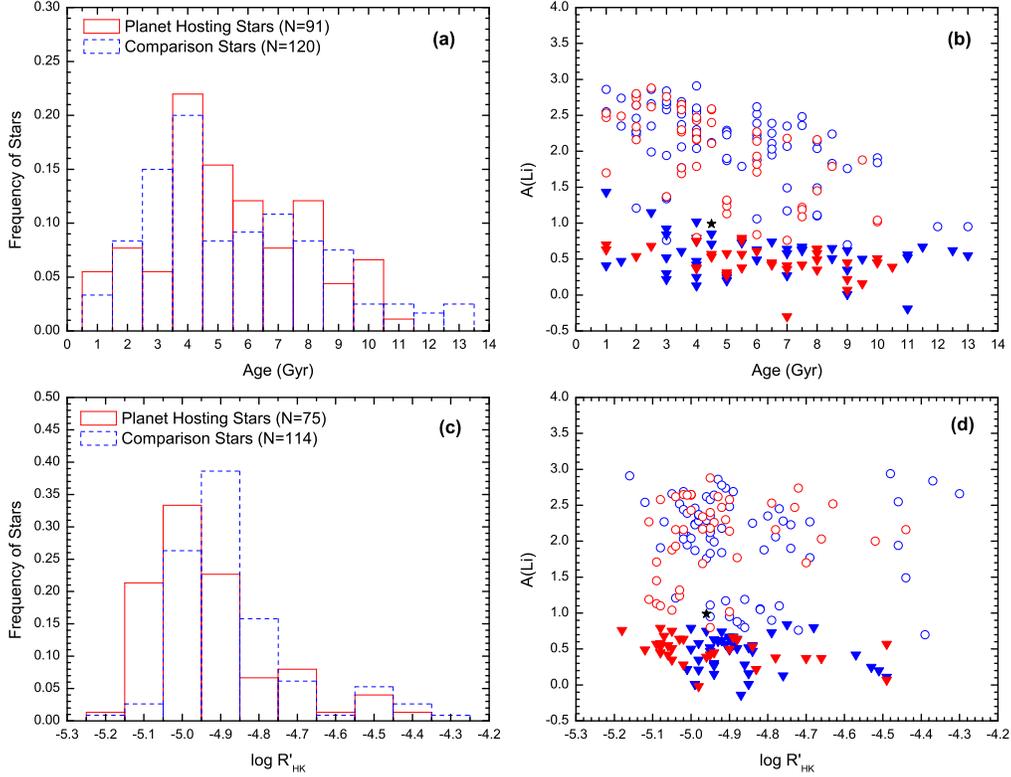}
\caption{(a) Age distributions of planet hosting (solid red line) and comparison (blue dashed line) 
stars. (b) Lithium abundances \textit{versus} age for stars with (red symbols) and without (blue symbols) 
planets. Open circles are detected lithium abundances and inverted triangles denote upper limit lithium abundances. 
The Sun is represented by a black star symbol. (c) Same as panel (a) but with the Ca II H+K emission index $\log R'_{HK}$. 
(d) Same as panel (b) but replacing age with $\log R'_{HK}$.}
\label{comp2}
\end{figure}

The age distributions for the sample stars are shown in panel \textquoteleft a\textquoteright\ of Figure \ref{comp2}. The stellar ages were taken from Table 4 
of \citetalias{ghezzi10a} (see that paper for details). An inspection of the two histograms in this panel (red solid line representing stars with planets
and blue dashed line representing stars without planets) indicates that overall there is not a significant difference between
the age distributions of the two samples; the average values are: $\langle Age\rangle_{P-H stars}$= 5.24 $\pm$ 2.48 Gyr and 
$\langle Age\rangle_{non P-H stars}$ = 5.41 $\pm$ 2.86 Gyr. 
The probability that these two samples belong to the same parent population (from a KS test) is 84\%. 
In panel \textquoteleft b\textquoteright\ of Figure \ref{comp2}, we observe an expected slight dependence of A(Li) 
upon age, with lower abundances being found for older stars. However, the behavior of stars with and without 
planets seems indistinguishable. 

Panel \textquoteleft c\textquoteright\ of Figure \ref{comp2} shows histograms comparing the chromospheric activity indices ($\log R'_{HK}$)
for the two samples studied here. 
The values of $\log R'_{HK}$ for the target stars were taken, whenever available, from \citet{h96}, \citet{t02} and \citet{w04}. 
An evaluation of possible systematic differences between the chromospheric activity indices in these
studies follows. \citet{t02} find that the activity indices in their study are consistent with those from \citet{h96}. 
A comparison of indices for 13 stars in our sample which appear in both studies finds good agreement:
the linear fit to the two datasets has a slope of 0.98 $\pm$ 0.12; a correlation coefficient R = 0.92 and a mean difference 
(in the sense \textquotedblleft Tinney et al. - Henry et al.\textquotedblright) of -0.01 dex. 
The works of \citet{h96} and \citet{w04} have 17 stars in common with our sample. 
A linear fit for these values reveals no significant trend, having a slope of 0.99 $\pm$ 0.04 and 
a correlation coefficient R = 0.99. There is just an offset in the averages of $\sim$0.04 dex, with the indices from 
\citet{h96} being higher than those in \citet{w04}. \citet{h96} derived an error of 0.052 for $\log R'_{HK}$, while \citet{w04} 
state that the uncertainties in $R'_{HK}$ are not larger than 13\% (which would correspond to $\sim$0.057 
for $\log R'_{HK} \simeq$ -4.90). 
As the offset between the two studies is compatible with the errors in $\log R'_{HK}$, no
corrections in the indices were atempted in order to bring them to a consistent scale. 
In this study, whenever a star had more than one $\log R'_{HK}$ value, the average value was adopted. 

Inspection of Panel \textquoteleft c\textquoteright\ of Figure \ref{comp2} indicates that there is a visible offset between the distributions of $\log R'_{HK}$ for stars with 
and without planets, with a probability of 6.80 $\times 10^{-5}$ that they are drawn from the same population. 
This result is a consequence of the greater number of planet hosting stars with low activity levels. We note 
that this fact may be related to a bias in the Doppler detection method, which is more accurate for inactive 
stars. In spite of this offset, the average chromospheric activity indices for planet hosting and comparison 
stars are, respectively, $\langle \log R'_{HK} \rangle$ = -4.94 $\pm$ 0.16 and -4.87 $\pm$ 0.17 dex. The overall global difference of 0.07 dex 
is interesting but cannot be considered as significant since it is roughly of the same order of the uncertainties 
discussed above. 
Panel \textquoteleft d\textquoteright\ of Figure \ref{comp2} shows 
that there is no clear correlation between Li abundances and $\log R'_{HK}$ and no visible difference in the 
behavior of stars with and without planets. 
Finally, the Sun ($\log R'_{HK}$ = -4.96; \citealt{w04}) does not seem to be anomalous.

The comparisons of the various quantities discussed above show that the samples of planet host and 
comparison stars are similar in their mass and age distributions, 
although the metallicities do show a significant difference, with the planet hosting stars being on average
more metal rich (as discussed in \citetalias{ghezzi10a}). Small differences are found in their respective
activity levels, but neither the metallicity nor the activity levels produce obvious trends with the lithium abundances. 


\subsection{Lithium Abundances}

\label{lithium}

\begin{figure}
\epsscale{1.00}
\plotone{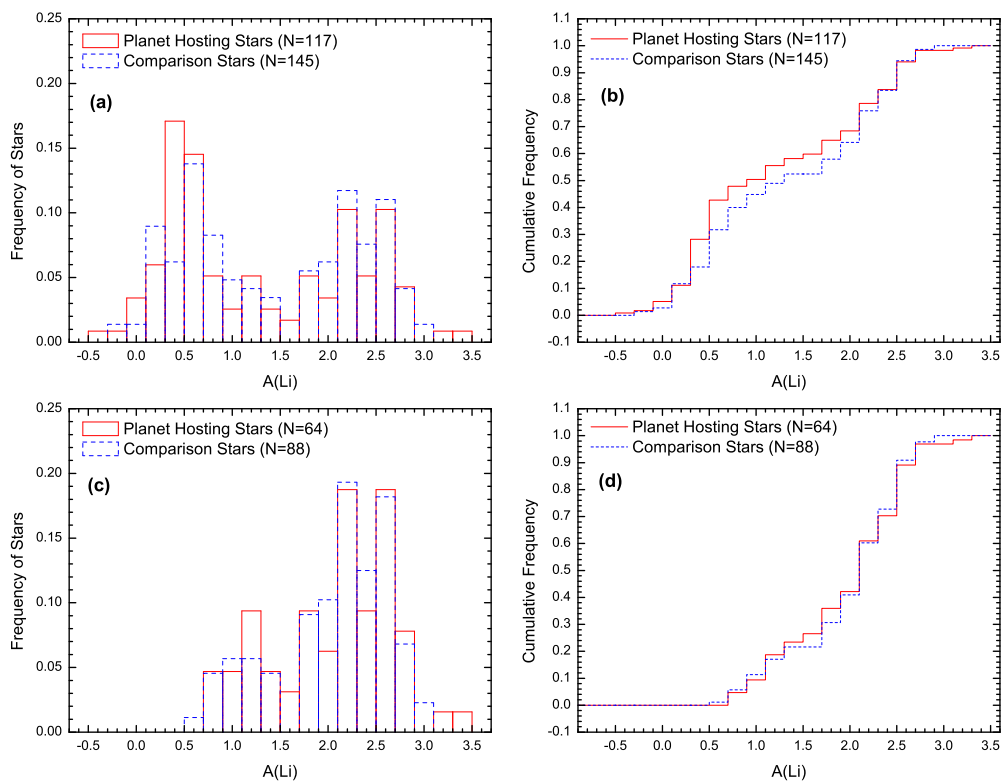}
\caption{Distributions of lithium abundances for planet hosting (red solid lines) and comparison (blue dashed 
lines) stars. Panels (a) and (b) show the frequency and cumulative distributions, respectively, for all stars 
in our sample. Panels (c) and (d) presents the same plots, but only for those stars with detected 
Li abundances.}
\label{li_hist}
\end{figure}

Possible differences between Li abundances in stars with and without planets remains
an active topic of discussion (\citealt{i09}; \citealt{s10}; \citealt{g10}; \citealt{b10}). 
In order to investigate such possible differences, we first examine the lithium abundance distributions 
for all stars in our sample, which are shown as the histograms in panel \textquoteleft a\textquoteright\ of Figure \ref{li_hist}. In this initial
discussion, upper limits were included in the construction of this plot. 
An inspection of panel \textquoteleft a\textquoteright\ reveals two interesting features. 
First, we can clearly see a bimodal distribution, with peaks located at A(Li) $\sim$ 
0.5 and 2.4 dex. Second, the similarity in the distributions of stars with and without planets is evident, 
with a significant difference being visible only in the bin centered at A(Li) = 0.4 dex, which is
dominated by upper limits; these facts become 
more evident if we analyze the cumulative distributions which are shown in panel \textquoteleft b\textquoteright\ of Figure \ref{li_hist}. 
The differences between the distributions are probably a result of the inclusion of those stars with only
upper limits to the Li abundances. In fact, if only detected Li abundances are considered, the 
distributions become more similar, as can be seen in panels \textquoteleft c\textquoteright\ (histogram) and \textquoteleft d\textquoteright\ (cumulative distribution)
of Figure \ref{li_hist}. The average detected lithium abundances of stars with and without planets are 
$\langle$ A(Li) $\rangle$ = 2.06$\pm$0.63 and 2.06$\pm$0.60 dex, respectively, and the probability 
that the two samples belong to the same parent population is 87\% (KS test). Therefore, this broad comparison 
of the entire sample does not seem to reveal any anomaly in the Li abundances of planet hosting stars.


\begin{figure}
\epsscale{1.00}
\plotone{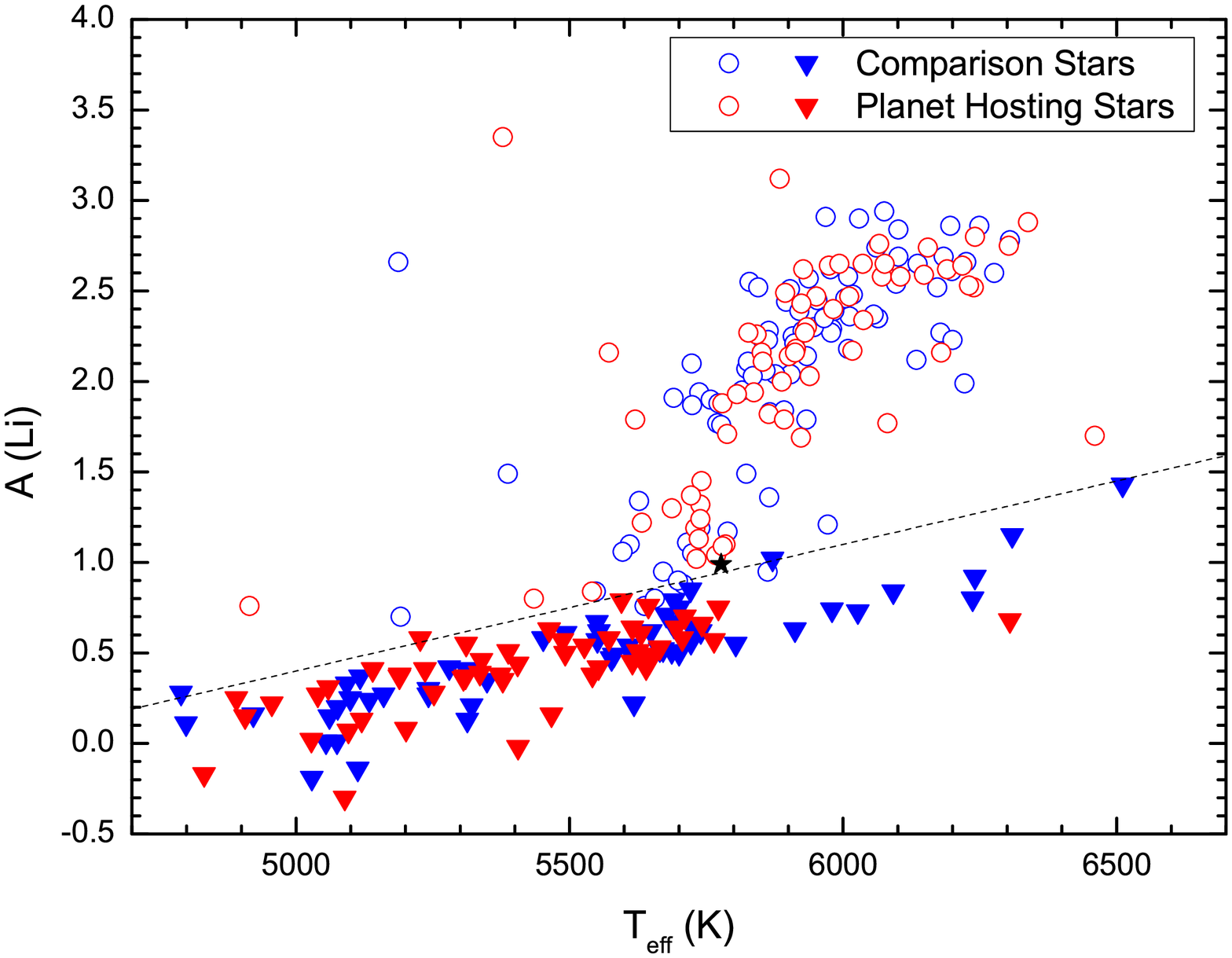}
\caption{Lithium abundances versus effective temperatures for planet hosting (red symbols) and comparison 
(blue symbols) stars. Determined abundances are represented by circles and upper limits are denoted by 
downard triangles. The dashed line shows the detection limit of our method. The black star represents the Sun.}
\label{li_teff}
\end{figure}

\begin{figure}
\epsscale{0.80}
\plotone{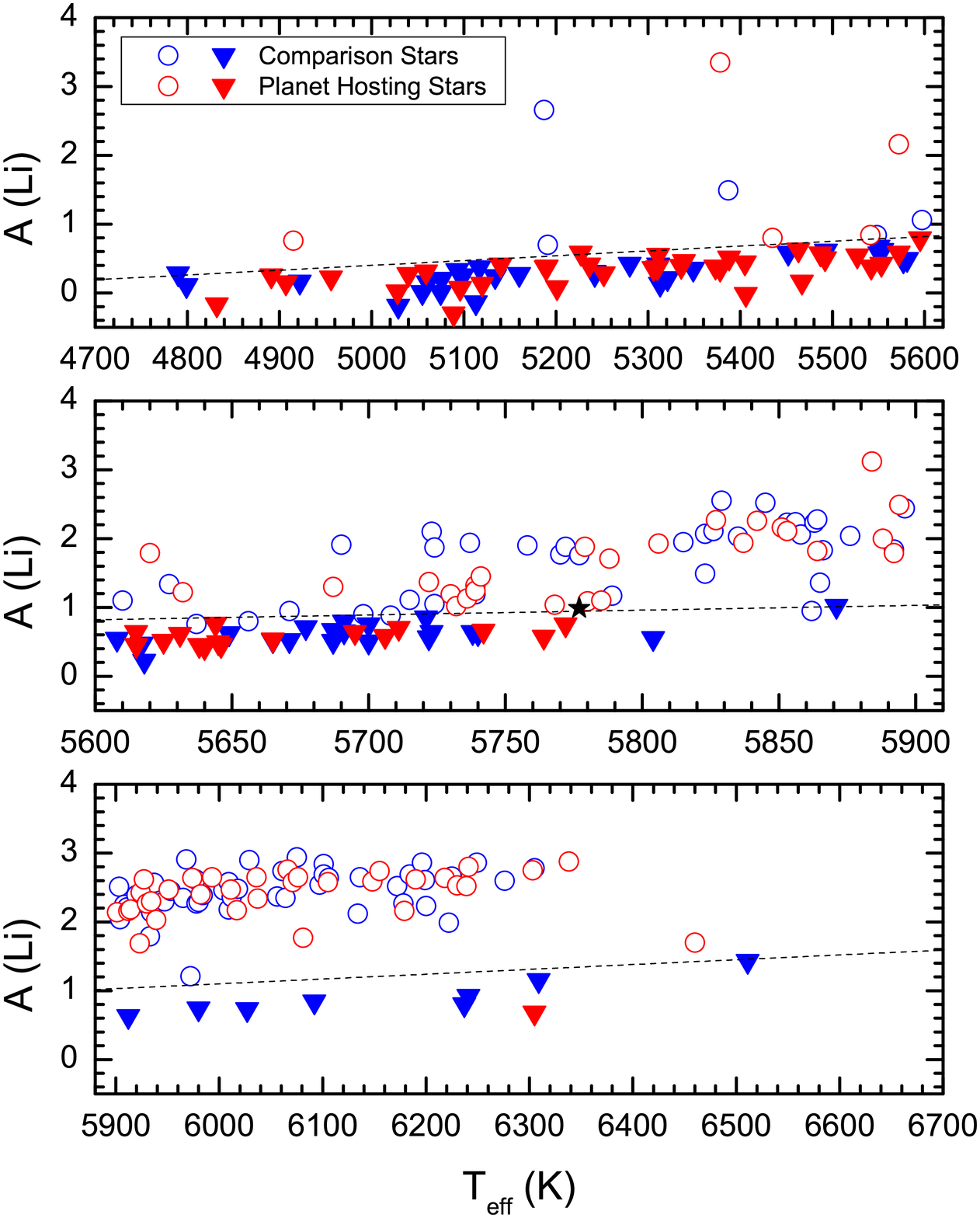}
\caption{Lithium abundances versus effective temperatures for the intervals $T_{eff} <$ 5600 K (upper 
panel), 5600 K $\leq T_{eff} \leq$ 5900 K (middle panel) and $T_{eff} >$ 5900 K (lower panel). 
The symbols and dashed line have the same meaning as in Figure \ref{li_teff}.}
\label{li_teff2}
\end{figure}

Differences between Li abundances in stars with and without planets have been reported in
the effective
temperature range 5600 K $\lesssim T_{eff} \lesssim$ 5900 K (\citealt{i09}).
In order to investigate this behavior in our samples, the lithium
abundances derived in this study are plotted versus the effective temperatures of the
target stars in Figure \ref{li_teff}. This plot holds some
interesting features. First, the detection limit of our method varies with T$_{eff}$
roughly as 0.07 dex/100 K, increasing from $\sim$0.3 dex at $\sim$4800 K to $\sim$1.5 dex at
$\sim$6500 K. Also, the well known decrease in lithium abundances with declining temperatures is recovered
(see, e.g., \citealt{tk05,lh06}). This behavior is ultimately a reflection of the trend seen in panel
\textquoteleft d\textquoteright\ of Figure \ref{comp1}, as
higher mass stars on the main sequence also have higher effective temperatures.
As there are three clearly distinct Li regimes in Figure \ref{li_teff} (hotter stars with high A(Li);
intermediate T$_{eff}$ stars with A(Li) declining and low T$_{eff}$ stars with Li upper limits), in the 
following we discuss them separately (see Figure \ref{li_teff2}).


For $T_{eff} <$ 5600 K (upper most panel of Figure \ref{li_teff2}), almost all stars have very low Li abundances 
($\lesssim$0.7 - 0.8 dex) and only upper limits could be derived in most cases. With only upper limits,
it is not possible to probe any possible differences in A(Li) in stars hosting planets compared to stars
not known to host planets. The exceptions to upper limits are: 
HD 1237, HIP 14810, AB Pic, HD 69830 and HD 181433 (stars with planets); HD 17925, HD 36435, HD 118972, 
HD 128674 and HD 212291 (stars without planets). The two stars exhibiting extremely high Li abundances are AB Pic (A(Li)=3.35 dex) 
and HD 17925 (A(Li)=2.66 dex). As already noted by \citet{tk05}, HD 17925 is a young star with high activity 
level ($\log R'_{HK}$ = -4.30; \citealt{h96}). 
AB Pic is a young star identified as a member of the 
Tucana-Holorogium association (\citealt{s03}), which has an estimated age of $\sim$30 Myr.
In \citetalias{ghezzi10a}, upper limits of 1 Gyr were derived for the ages of these two stars. The same age upper limit is derived 
in \citetalias{ghezzi10a} for HD 1237, 
which has A(Li) = 2.16 dex and is also a young, active star (age of 0.02 Gyr and $\log R'_{HK}$ = -4.27; 
\citealt{n01}). The stars HD 36435 and HD 181433 have uncertain ages because they are located in 
crowded regions of their respective grids of isochrones. Based on its chromospheric activity level 
($\log R'_{HK}$ = -4.44; \citealt{h96}), HD 36435 is most likely young, which would explain its relatively 
high Li abundance (A(Li)= 1.49 dex). HD 181433, on the other hand, has $\log R'_{HK}$ = -5.11 (\citealt{b09}) and does 
not seem to be young. Although consistent with the chromospheric activity level, its lithium abundance 
(A(Li)= 0.76 dex) is relatively high when compared to stars of the same temperature (difference of $\sim$ 
0.35 dex). The lithium abundance of HD 212291 also seems to be higher (by $\sim$0.25 dex) than thetypical value for stars with similar T$_{eff}$. 
Although its age is uncertain (the star is located in a crowded region of its 
grid of isochrones), its activity level ($\log R'_{HK}$ = -4.82; \citealt{w04}) shows that it may be younger than the Sun, which could explain the value of the Li abundance.
Finally, the four remaining stars (HIP 14810, HD 69830, HD 118972 and HD 128674) have Li abundances 
consistent with those of other stars within the same temperature range. 
In this lower temperature regime ($T_{eff} <$ 5600 K), there are 41 and 34 stars with and without planets, respectively; 
this corresponds to the largest sample analyzed to date.
If we do not consider the stars with high Li abundances 
(HD 1237 and AB Pic in the planet-hosting sample; HD 17925 and HD 36435 in the comparison sample), a general overlap of the two samples is clear. Taking 
the upper limits as real detections, we find that the average Li abundances of planet hosting and comparison 
stars are, respectively, $\langle A(Li) \rangle$= 0.37 $\pm$ 0.25 and 0.35 $\pm$ 0.28 dex. 

In the hotter main sequence stars in this sample, where
$T_{eff} >$ 5900 K (bottom most panel of Figure \ref{li_teff2}), most have A(Li) $\sim$ 1.5 - 3.0.
There are, however, a small number of stars with only upper limits, and we note 
that some of these stars (with low Li abundance 
and $T_{eff} \gtrsim$ 6300 K) may be in the Li dip (see \citealt{bt86,b95}). In this temperature interval, 
there are 34 and 54 planet hosting and comparison stars, respectively, with considerable overlap.
It is clear that the number of stars with planets diminishes with increasing temperature, reflecting the 
limitation of radial velocity surveys. Also, we notice that 8 comparison stars belong to the low-Li group, 
while only one star with a planet has an upper limit on its Li abundance. Although this feature can also be 
seen in \citet{tk05} and \citet{lh06}, we are not sure if this is related to any physical effect or selection 
bias. Considering only detected lithium abundances, we obtain $\langle A(Li)\rangle = 2.42 \pm 0.31$ 
and $2.44 \pm 0.32$ for stars with and without planets, respectively. 
Thus, we do not find the excess Li abundances for planet hosting stars with $T_{eff} \sim$ 6000 K
discussed in \citet{g08} and \citet{g10}, and find no detectable differences in the Li abundances
between stars with and without planets in the most massive stars of the sample (M $\gtrsim$ 1.2 M$_{\sun}$).

A more complicated behavior of the Li abundance occurs in the
transition region between high and low Li, within the temperature interval 5600 K $\leq T_{eff} 
\leq$ 5900 K (middle panel of Figure \ref{li_teff2}). In this temperature regime, there are 42 and 57 stars with and 
without planets, respectively in our sample. The overlap is considerable, except in the narrow range of $T_{eff}\sim$ 
5700 - 5800 K. 
Over the entire temperature range shown in this panel ($T_{eff}$ = 5600 - 5900 K), 26 planet hosting stars (or 62\%) 
have detected Li abundances, with an average value 
of $\langle A(Li)\rangle = 1.68 \pm 0.53$ dex. For the comparison stars, the average lithium abundance for 
37 stars (65\%) with detections is $\langle A(Li)\rangle = 1.69 \pm 0.54$ dex. Treating the upper limits as 
actual detections, we derive $\langle A(Li)\rangle = 0.58 \pm 0.11$ and $0.62 \pm 0.17$ dex for stars with and 
without planets. Finally, it is interesting to note that the Sun does not exhibit an excessive Li depletion; 
actually, it looks like a \textquotedblleft normal\textquotedblright\ main-sequence star.


Within this complex temperature transition region for Li, if we now follow the analysis of \citet{i09} and 
keep only stars within the very narrow range of $T_{eff}= 5777 \pm 80$ K, 
the sample here is left with 21 planet hosting and 27 comparison stars. 
For the sample of stars with planets, the average is $\langle A(Li)\rangle = 1.50 \pm 0.44$ dex for the 16 detections (76\%) and 
$\langle A(Li)\rangle = 0.65 \pm 0.08$ dex for the 5 upper limits. For the control sample, the average is
$\langle A(Li)\rangle = 1.76 \pm 0.48$ dex for the 19 detections (70\%) and $\langle A(Li)\rangle = 0.64 \pm 0.11$ 
dex for the 8 upper limits. Following the analysis of \citet{g10}, on the other hand, we have 20 and 30 planet 
hosting and comparison stars, respectively, in the temperature interval 5650 - 5800 K. For the former group, 
we obtain $\langle A(Li)\rangle = 1.30 \pm 0.26$ dex for the 13 detections (65\%) and 
$\langle A(Li)\rangle = 0.63 \pm 0.08$ dex for the 7 upper limits. For the control sample, we derive 
$\langle A(Li)\rangle = 1.45 \pm 0.47$ dex for the 16 detections (53\%) and $\langle A(Li)\rangle = 0.64 \pm 0.11$ dex for the 14 upper limits.
These comparisons between Li abundances in planet hosting stars and comparison stars using
the different T$_{eff}$ boundaries as defined by \citet{i09} and \citet{g10}
are intriguing, with averages differences of +0.26 dex and +0.15 dex, respectively.
Note, however, that in the latter case, in particular, the average lithium abundances in the
two samples overlap within our estimated uncertainties.

It is interesting to investigate
in more detail the Li abundances derived here for stars falling
in the narrow T$_{eff}$ range centered on the solar effective
temperature as used by \citet{i09}, since this 
temperature interval resulted here in the larger difference of +0.26
dex. The Israelian et al. sample in this T$_{eff}$ interval
contains more stars than analyzed here, with 23 planet hosting
stars and 60 comparison stars, compared to 21 and 27 stars here.
The Li abundance distribution from \citet{i09} is 
heavily influenced by non-detections of \ion{Li}{1} and thus, upper
limits to A(Li), with 17 and 28 upper limits for planet hosting
stars and comparison stars, respectively. Taking the remaining
\ion{Li}{1} detections from Israelian et al., which span values of 
A(Li) $\sim$ 1.0 - 3.0, and examining the cumulative fraction of
A(Li) reveals a shift towards larger Li abundances in the stars
without planets, as discussed in \citet{i09}. This same procedure is carried out for
the samples of stars here, where there are only 5 upper limits in planet
hosting stars and 8 upper limits in comparison stars. The
cumluative fractions of A(Li) in the two sets of stars also show
that the comparison stars are shifted towards somewhat larger
values of A(Li), but not as large a shift as found by \citet{i09}. 
Taken together, we find the same effect as found
by Israelian et al., however the difference in the samples here is
not as large. It is worthwhile to probe for other differences
between planet hosting and comparison stars that might pertain
to these possibly real differences in Li abundances between the
two sets of stars.

Stellar rotation would be an obvious candidate on which to focus
more closely in trying to understand why there might be differences
in A(Li) between the planet hosting and comparison star samples, 
however such a study is not feasible at the spectral resolution
of the observed spectra studied here given the considerable degeneracy
between macroturbulence and $v \sin i$.

\begin{figure}
\epsscale{1.00}
\plotone{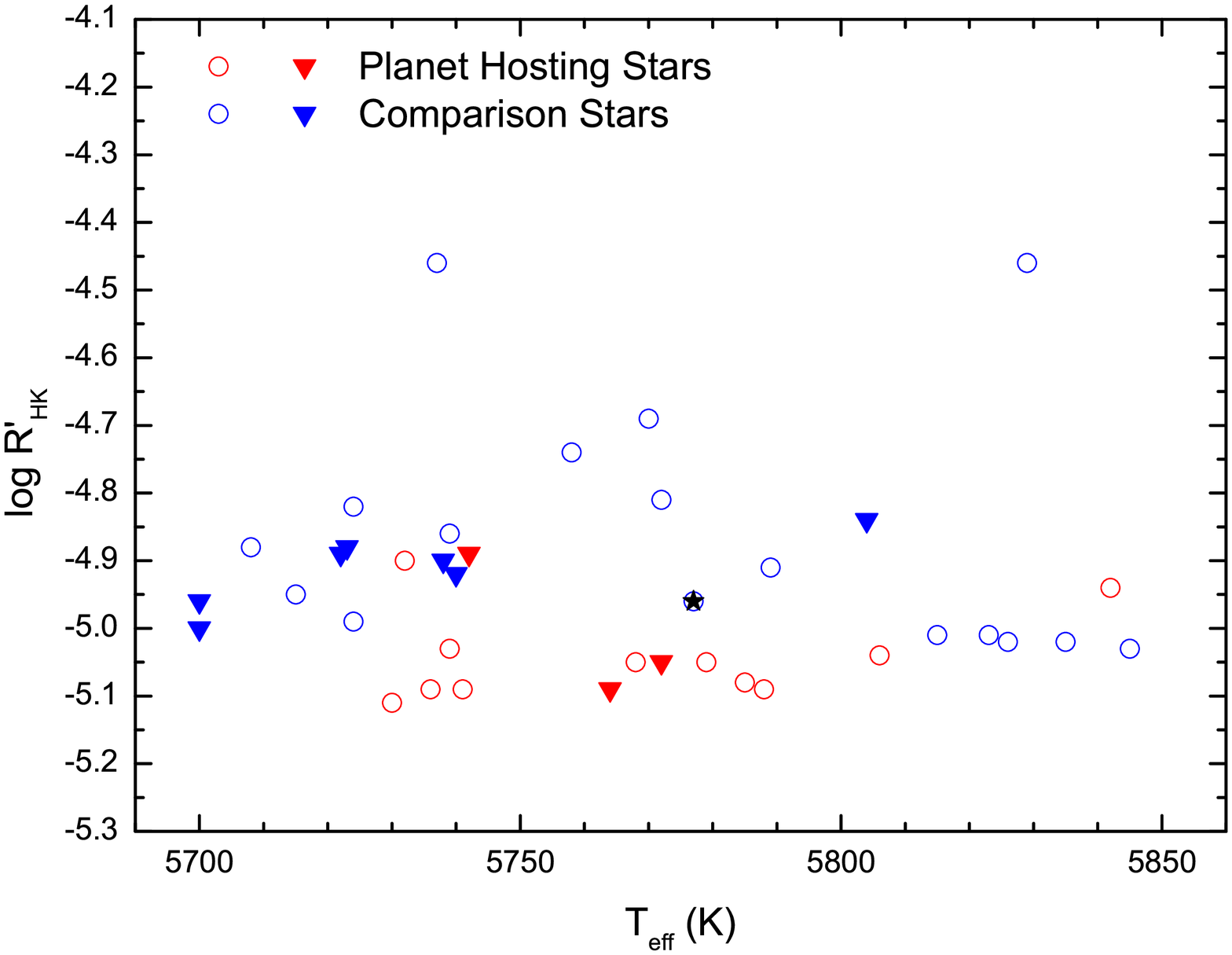}
\caption{Ca II H+K emission indices versus the effective temperatures for planet hosting (red symbols) and comparison (blue symbols) stars in the effective temperature between 5700 -- 5850 K. Determined abundances are represented by circles and upper limits are denoted by 
downard triangles. The black star represents the Sun.}
\label{logRhk_teff_sun}
\end{figure}

Another observational variable to investigate is the stellar activity
level in the different types of stars studied here for lithium. The 
index of stellar activity discussed in Section \ref{comp} is $\log R'_{HK}$
and this index is plotted in Figure \ref{logRhk_teff_sun} versus T$_{\rm eff}$ over the 
limited range 5700 -- 5850 K. Within this narrow temperature range,
which was the focus of the discussion in \citet{i09}, it is
clear that the planet hosting stars tend to fall towards lower values of 
$\log R'_{HK}$ relative to the comparison stars. 


\begin{figure}
\epsscale{0.80}
\plotone{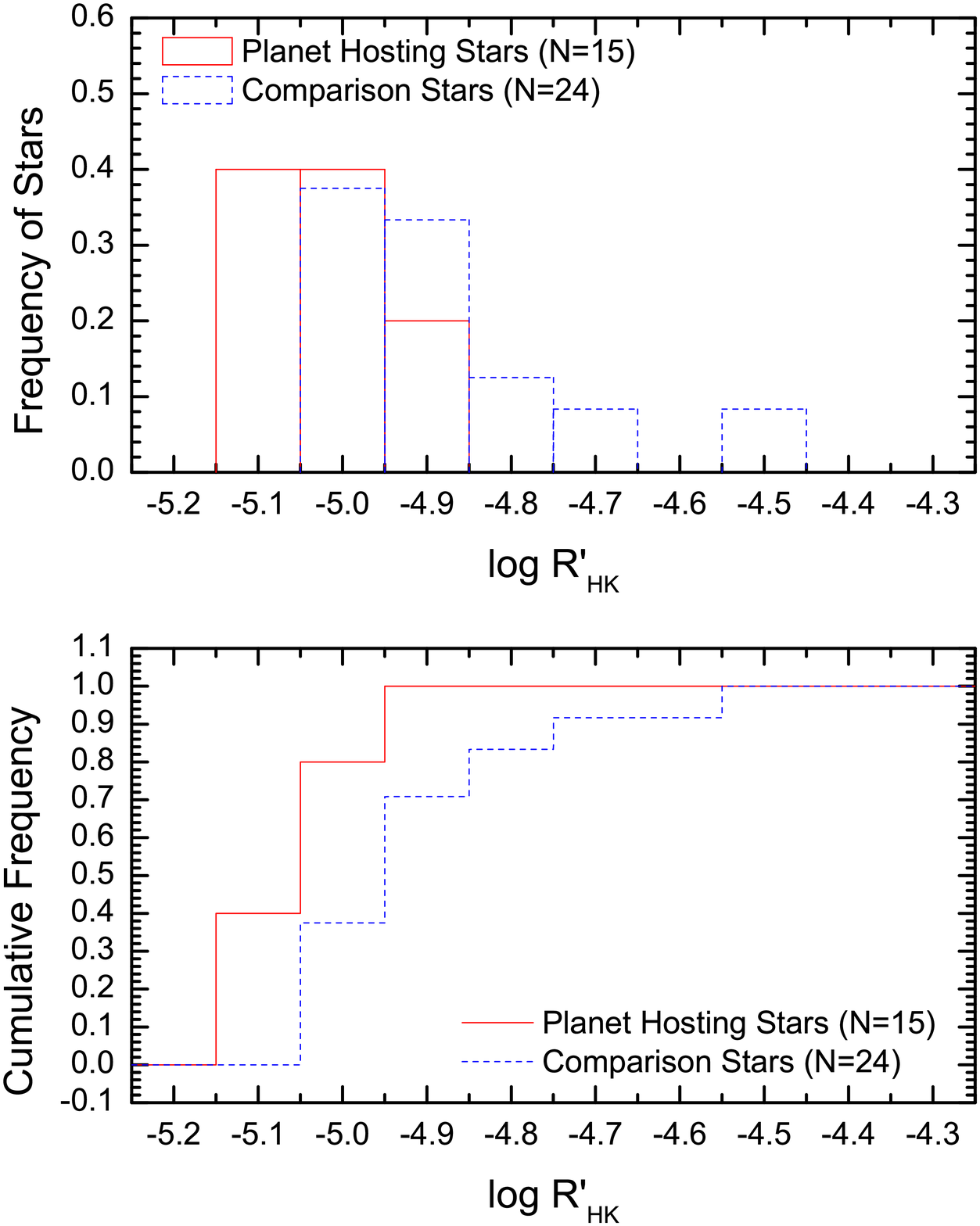}
\caption{Frequency (upper panel) and cumulative (lower panel) distributions of the Ca II H+K activity indices for planet hosting (red solid line) and comparison 
(blue dashed line) stars in the range between $T_{eff}$ = 5700 -- 5850 K.}
\label{logRhk_hist}
\end{figure}

The point noted above about differences in stellar activity levels 
is made clearer 
by the respective frequency distributions and cumulative fractions
of $\log R'_{HK}$ shown in the top and bottom panels of Figure \ref{logRhk_hist},
respectively. The comparison stars are shifted significantly towards
larger levels of stellar activity when compared to planet hosting stars.
Within a sample of stars which have a restricted range in mass and
effective temperature, such as the small subset of stars being 
discussed here, the value of $\log R'_{HK}$ is likely related to both
rotational velocity and age, with both the rotational velocities and
chromospheric activity decreasing with increasing age. As noted above,
all of the stars studied here within this range of T$_{\rm eff}$ = 5700 -- 5850 K 
are slow rotators, thus it is not possible to disentangle
directly differences in rotation between stars with and without planets.
The other main parameter associated with chromspheric activity and
rotation is age; a detailed comparison of derived ages from \citetalias{ghezzi10a} between
the planet hosting and comparison stars reveals no significant difference
in the age distributions, however it must be noted that the age estimates
come with large uncertainties of typically $\pm$2 -- 3 Gyr.


\citet{t10} has already noted the difference in chromospheric
activity levels between stars with and without planets and argue
that the differences in activity levels are rooted in differences in 
the underlying rotational velocities. The picture put forth in 
\citet{t10} is that the stars with planets tend to rotate more
slowly (with less stellar activity) due to the influence of giant planets,
or massive protoplanetary disks slowing down stellar rotation early in
the evolution of the parent star.

\citet{s10} have also investigated the behavior of the Li abundance
in stars with and without planets across the 5700 -- 5850 K range in
T$_{\rm eff}$ and find differences in the behavior of A(Li), with the
stars with planets having smaller abundances of Li. They investigate
the stellar parameters age and mass and conclude that neither the age
distributions nor the mass distributions are different for stars with
and without planets, and thus suggest that the observed difference in
lithium abundances are likely due to some process related to the formation
of planetary systems (as also concluded by \citealt{t10}).

Most recently \citet{b10}, in a study of 117 solar-like stars,
find strong correlations of a decrease in the lithium abundance with
stellar age, but do not find any differences in the behavior of Li in
planet hosting stars compared to stars without known planets when
rather restricted ranges in stellar masses and metallicities
are enforced. They
suggest that the previously identified differences were due to systematic
biases caused by combinations of age and metallicity in the samples.

\begin{figure}
\epsscale{1.00}
\plotone{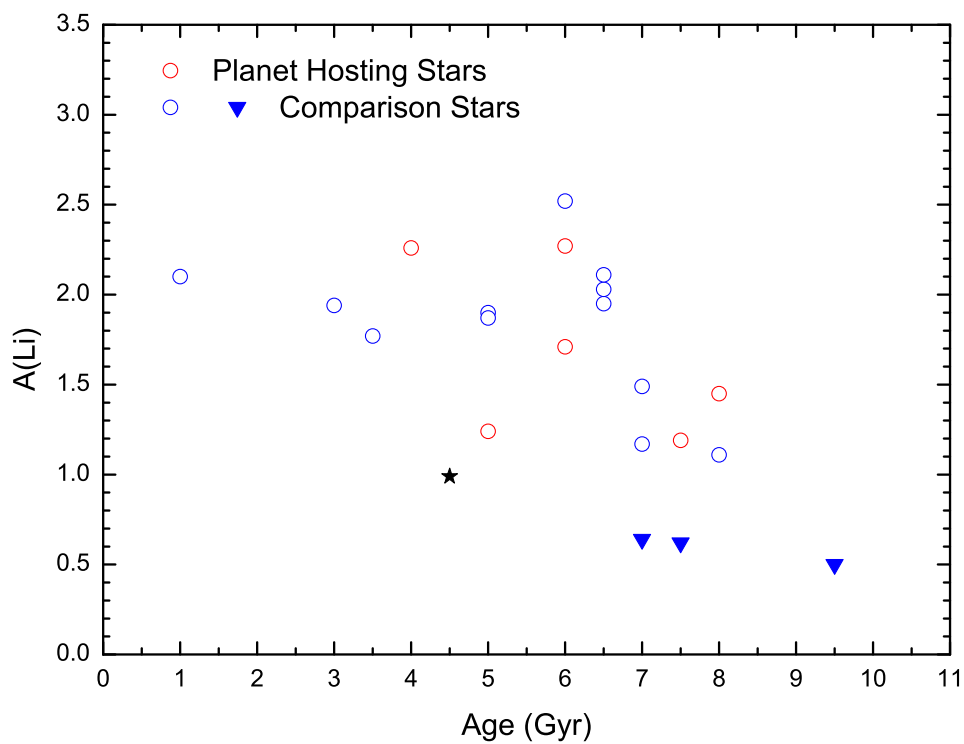}
\caption{Lithium abundances versus stellar ages for planet hosting (red symbols) and comparison 
(blue symbols) stars with $T_{eff}$ = 5700 -- 5850 K and [Fe/H] = 0.00 to +0.20 dex. 
The mean masses and mass ranges are very similar for these samples with 
the $\langle M\rangle_{P-H stars}$ = 1.12 $\pm$ 0.05 M$_{\sun}$ and $\langle M\rangle_{NonP-H stars}$ = 1.09 $\pm$ 0.06 M$_{\sun}$.
Measured abundances are represented by circles and upper limits are denoted by 
downard triangles. The black star represents the Sun.}
\label{li_age}
\end{figure}

We investigate the conclusions from \citet{b10} by restricting
the metallicity and mass of the sample of stars studied here and comparing
the A(Li) -- stellar age relations for stars with planets and those
stars without.  Using a similar procedure as that in \citet{b10},
we isolate a subset of stars from the already restricted range in
effective temperature (T$_{\rm eff}$= 5700 -- 5850 K) with [Fe/H] = 0.00
to +0.20 (or +0.10$\pm$0.10).  Within this parameter range, there are
12 stars without planets having measured Li abundances (and an additional
3 with upper limits), and 6 planet-hosting stars. The parameters of these
two sets are closely matched, with the means and standard deviations
of stars with planets and stars without planets being, respectively,
[Fe/H] = +0.12$\pm$0.05 and +0.10$\pm$0.03; M = 1.12$\pm$0.05 
and 1.09$\pm$0.06 M$_{\sun}$; and Age = 6.1$\pm$1.5 and 5.4$\pm$2.0 Gyr.
The values of A(Li) versus stellar age are plotted in Figure \ref{li_age} for the
planet hosting stars and stars without known planets.  In this case, there
is no significant difference in the behavior of A(Li) with stellar age
between the two stellar samples.  As noted by \citet{b10},
when the stellar parameters which are known to influence Li abundances
over time are restricted, the behaviors of A(Li) in stars with and
without planets appear very similar. 


Given that there remains some uncertainty in the detailed behavior of
A(Li) and the underlying processes that shape this behavior, the observation
noted here that stars with planets tend to exhibit lower stellar activity
levels remains. However, at this point, this signature cannot be separated from
observational biases which are present in planet detection from radial velocity measurements.
Lower values of stellar activity may result in smaller intrinsic
radial velocity variations, making the chromospherically quiet stars
targets around which planets producing small RV amplitudes can still
be detected.


\section{Conclusions}

\label{conc}

This paper is the third in a series which analyzes chemical abundances in a sample of planet
hosting stars and a comparison sample of disk stars. 
\citetalias{ghezzi10a} (\citealt{ghezzi10a}) presented metallicities, [Fe/H], for dwarf stars; Paper II (\citealt{ghezzi10b}) 
analyzed samples of subgiants and giant stars and the present paper analyzes the element lithium in the same main-sequence targets
as in \citetalias{ghezzi10a}. These targets consist of 117 planet hosting stars and 145 comparison stars known not to host
giant planets. 

Lithium abundances are potentially important for a better understanding of processes involved 
in the formation and evolution of planetary systems. 
Several recent studies in the literature have
looked into possible differences between Li abundances of stars with and without planets,
but so far have yet to reach a full consensus regarding a possible excess Li depletion in planet hosting 
stars (see e.g. reviews by \citealt{m10} and \citealt{santos10}). 
The combination of uniform data and homogeneous analysis with well-selected
samples makes the results in this study well-suited to investigate 
possible differences in the Li abundances found in planet hosting stars.

Lithium abundances were derived 
through an automated profile fitting of the \ion{Li}{1} resonance doublet at 
$\lambda$6707.8 \AA. 
As a test, the automated analysis of the Sun as a normal target star yields a lithium abundance A(Li) = 0.99 $\pm$ 0.16; this result
is in good agreement with recent NLTE 3D abundance determinations done by \citet{a09} and
\citet{c10}. 
When compared to stars with similar properties in our sample (irrespective of the presence of planets), 
the Sun appears to have a normal Li abundance.

Lithium abundances were not found to exhibit significant trends with metallicity and stellar activity levels. It is possible, though, 
to identify a slight negative correlation with age. The most significant dependance is with effective temperature 
or, alternatively, mass. Hotter (more massive) stars have higher Li abundances. This is a well known effect 
related to the variation of the depth of the convective zone as a function of mass.

The most suggestive difference between planet hosting stars and the comparison stars occurs in the narrow
T$_{eff}$ range from 5700 to 5850 K, where the behavior of the Li abundance is perhaps the most
complex. Within this restricted temperature range, we find that the comparison stars tend to have
on average somewhat larger Li abundances than the planet hosting stars; this is similar to earlier results from
\citet{i09} and \citet{g10}, although the differences found here are smaller.
Within this sample of stars, the planet hosting stars also tend to exhibit lower
levels of stellar chromospheric activity $\log R'_{HK}$ relative to the comparison stars,
thus mirroring the behavior of the lithium abundances.
This difference in chromospheric activity could be due to planet-hosting
stars being somewhat older or rotating more slowly (or a combination of
both). Perhaps a more likely explanation is that this is a selection effect,
as the chromospherically quieter stars exhibit smaller radial velocity jitter, and
thus represent easier stars around which to find Doppler-detected planets.

Within the effective temperature interval between 5700 to 5850 K, when more constrained samples of stars having quite
restricted ranges in stellar mass and metallicity are compared, however,
no differences in the behavior of A(Li) versus stellar age were found between
stars with and without (known) planets.
The results for lithium abundances presented here highlight the
importance of comparing samples which are as restricted as possible in
all parameters, however, in such cases conclusions are based on a much smaller
number of objects and thus require further confirmation.
As the number of planet hosting stars continues to grow, larger samples with restricted
masses and metallicities will be available to probe the real behavior of lithium
and to fully understand the influence of planets on the lithium abundance of
the host stars.


\acknowledgements{We thank the referee Martin Asplund for suggestions which helped to improve the paper.
We thank Simon Schuler for discussions. L.G. acknowledges the financial support of CNPq.
Research presented here was supported in part by NASA grant NNH08AJ581.}


\begin{deluxetable}{lccc}
\tablecolumns{4}
\tablewidth{0pt}
\tablecaption{Lithium abundances.\label{li}}
\tablehead{
\colhead{Star} & \colhead{FWHM$_{Gauss}$} & \colhead{A(Li)} & \colhead{$\delta$A(Li)} \\
\colhead{} & \colhead{(\AA)} & \colhead{} & \colhead{}}
\startdata
\sidehead{\textit{Planet Hosting Stars}}
HD 142  & 0.379 & 2.88 $\pm$ 0.03 & 0.11 \\
HD 1237 & 0.208 & 2.16 $\pm$ 0.02 & 0.10 \\
HD 2039 & 0.164 & 2.30 $\pm$ 0.03 & 0.11 \\
HD 2638 & 0.140 & $\leq$0.41 & \nodata \\
HD 3651 & 0.155 & $\leq$0.28 & \nodata \\
\sidehead{\textit{Comparison Sample}}
HD 1581 & 0.211 & 2.25 $\pm$ 0.05 & 0.11 \\
HD 1835 & 0.237 & 2.55 $\pm$ 0.02 & 0.10 \\
HD 3823 & 0.165 & 2.36 $\pm$ 0.02 & 0.10 \\
HD 4628 & 0.175 & $\leq$0.01 & \nodata \\
HD 7199 & 0.140 & $\leq$0.35 & \nodata \\
\enddata
\tablecomments{Table \ref{li} is published in its entirety in the eletronic edition of the Astrophysical 
               Journal. A portion is show here for guidance regarding its form and content.}
\end{deluxetable}


\end{document}